\documentclass[conference]{IEEEtran}
\IEEEoverridecommandlockouts
\usepackage{cite}
\usepackage{gensymb}
\usepackage{amsmath,amssymb,amsfonts}
\usepackage{algorithmic}
\usepackage{graphicx}
\usepackage{soul}
\usepackage{textcomp}
\usepackage{xcolor}
\def\BibTeX{{\rm B\kern-.05em{\sc i\kern-.025em b}\kern-.08em
    T\kern-.1667em\lower.7ex\hbox{E}\kern-.125emX}}
\begin{document}

\title{Autoencoder-based Anomaly Detection \\in Smart Farming Ecosystem}

\author{\IEEEauthorblockN{Mary Adkisson\IEEEauthorrefmark{1}, Jeffrey C Kimmell\IEEEauthorrefmark{2}, Maanak Gupta\IEEEauthorrefmark{3}, and Mahmoud Abdelsalam\IEEEauthorrefmark{4}}
\IEEEauthorblockA{\IEEEauthorrefmark{1}\IEEEauthorrefmark{2}\IEEEauthorrefmark{3}{Dept. of Computer Science},
{Tennessee Technological University},
Cookeville, Tennessee 38505, USA \\\IEEEauthorrefmark{4}Dept. of Computer Science, North Carolina A\&T State University,
Greensboro, NC, USA\\}
\IEEEauthorrefmark{1}maadkisson42@tntech.edu,
\IEEEauthorrefmark{2}jckimmell42@tntech.edu, 
\IEEEauthorrefmark{3}mgupta@tntech.edu
\IEEEauthorrefmark{4}mabdelsalam01@ncat.edu}

\maketitle

\begin{abstract}
The inclusion of Internet of Things (IoT) devices is growing rapidly in all application domains. Smart Farming supports devices connected, and with the support of Internet, cloud or edge computing infrastructure provide remote control of watering and fertilization, real time monitoring of farm conditions, and provide solutions to more sustainable practices. This could involve using irrigation systems only when the detected soil moisture level is low or stop when the plant reaches a sufficient level of soil moisture content. These improvements to efficiency and ease of use come with added risks to security and privacy. 
Cyber attacks in large coordinated manner can disrupt economy of agriculture-dependent nations. 
To the sensors in the system, an attack may appear as anomalous behaviour. In this context, there are possibilities of anomalies generated due to faulty hardware, issues in network connectivity (if present), or simply abrupt changes to the environment due to weather, human accident, or other unforeseen circumstances.
To make such systems more secure, it is imperative to detect such data discrepancies, and trigger appropriate mitigation mechanisms. 


In this paper, we propose an anomaly detection model for Smart Farming using an unsupervised Autoencoder machine learning model. We chose to use an Autoencoder because it encodes and decodes data and attempts to ignore outliers. When it encounters anomalous data the result will be a high reconstruction loss value, signaling that this data was not like the rest. Our model was trained and tested on data collected from our designed greenhouse test-bed. Proposed Autoencoder model based anomaly detection achieved 98.98\% and took 262 seconds to train and has a detection time of .0585 seconds.


\end{abstract}

\section{Introduction}
\begin{figure}[t!]
    \centering
    \includegraphics[width=\columnwidth]{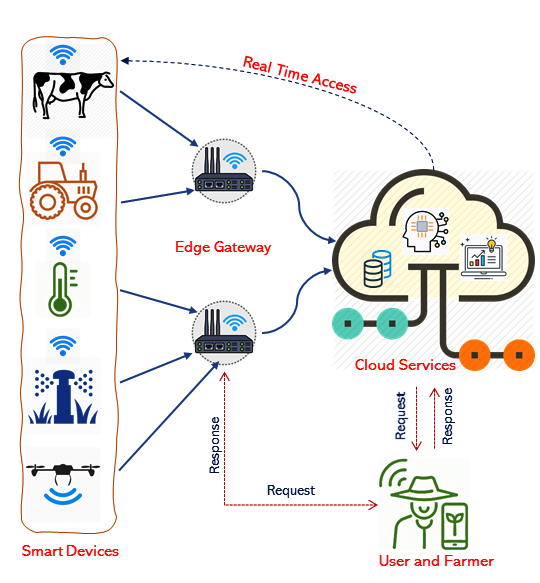}
    \caption{Smart Farming Conceptual Architecture \cite{gupta2020security}.}
    \label{fig:smart_farming_interaction}
\end{figure}

Smart farming is the implementation of IoT technology in a traditional farm environment. Farms provide food, jobs, and commerce across the globe. The addition of technology to this system has the potential to reduce soil depletion by monitoring crop growth patterns and reduce the amount of fertilizer and water used by optimizing schedules for each to match weather patterns and specific crop needs. Smart farms also have the potential to improve crop yields, as well as increase levels of sustainability \cite{iotfood}. Figure \ref{fig:smart_farming_interaction}, shows an end to end interaction among various entities involved in the smart farming ecosystem. The result of a successful smart farm would be decreased waste and increased output, all while making the process easier for the farmer. However, farming is a particularly critical sector due to the world's dependency on its physical output. A disruption in food supply would have negative consequences on even a small farm and the people who depend on its output. These consequences and dependence become greater the larger the farm is. Disruptions could come in the way of device failure, natural disaster, or attack on the system.  

There are different IoT devices that can be used within a smart farming system including, but not limited to, sensors for soil moisture level, temperature, humidity, etc., actuators to control light level, air circulation, watering, fertilizer, and many more. Smart farming devices are often exposed to harsh conditions such as extreme heat and light, as well as condensation build up or even flooding. Although the price of IoT devices is low, making them affordable for any level of farmer to use, they have inherent flaws and limitations. Mostly have minimal or no security protocols and overall low cost of hardware parts. This means that these devices are easy to replace, but also sacrifice consistency in readings, and as mentioned before, little or no privacy and security mechanisms. IoT devices are highly susceptible to failure and manipulation by attackers. An example of an attack could be for an adversary to target a smart farming infrastructure to disrupt food production simply to cause harm or to gain financially by placing holding the systems hostage and demanding a ransom before relinquishing control. Simply by deploying IoT devices for smart farming purposes would inherit different types of security risks that the farmers and community previously would not have to worry about. Further, with the exponential rise in the number of IoT devices in the world has introduced new types/variations or degree of risks in security and privacy. These devices often have unsatisfactory security practices such as weak/guessable passwords, insecure network services, etc. 

A list of 10 of the biggest areas of insecurity in IoT devices can be found in the OWASP IoT Top 10 document \cite{owasp}. To make IoT devices' security flaws even worse, these devices are also deployed on a massive scale, which means that the vulnerability of a single sensor could be exacerbated by using 10s or 100s of the same sensor on a large farm \cite{tawalbeh2020iot}. IoT devices often have Bluetooth, WiFi, or other network connectivity capabilities. This is what makes these devices captivating and innovative. They allow the user to monitor farm or greenhouse conditions remotely and, in advanced scenarios, control actuation to ensure that optimal conditions are upheld. However, network connectivity offers hackers a pathway to perform attacks and directly opens a farm up to the dangers of the Internet. Apparently the aspects of smart farming that make it beneficial, internet connectivity and use of cheap IoT devices are also what put the system at risk. This is because IoT devices make it possible for cyber attacks to move past cyberspace and into the physical world \cite{fu2021hawatcher}. In addition, in connected systems like smart farming where data reading can result in actuation of other devices, it is critical to identify anomalous behaviour timely. 

In this paper, we focus on detecting anomalous behaviour of different sensors deployed in a smart farming ecosystem. Determining anomaly in this critical and time sensitive domain is imperative to curtail and limit the cyber risk and provide an opportunity for activating/deploying relevant security mitigation solutions. In addition, some sensor readings result in action from other devices, for example, a low moisture reading will result in actuate a water sprinkler. Therefore, it is critical to timely identify even a slightest anomalous behaviour (which could be because of a faulty sensor, or cyber attack) to prevent large scale damage to ecosystem. Our goal is to implement machine learning to accurately and quickly detect anomalies that could be the result of device failure, accidental interference, or by attacks. We envision that by exploring anomalous behavior and mitigation techniques, we can make smart farming safer to use and expand the work being done in this domain. The entire overview of our approach is as follows:
The first step was to select sensors, set up a test environment, and collect data. The data we collected from sensors deployed in a greenhouse were independent from any network connectivity. This was done to focus on collecting data that represents normal, healthy conditions within the greenhouse. Although we do not study the effects of potential attacks or anomalies due to network connectivity, our model would still be able to detect anomalies because we trained solely on non-anomalous data. At this point, deployment in commercial farm settings is beyond the scope of our research, and we are only focused on developing a model which can detect anomalies. Therefore, we leave the possibility of connectivity up to future work. After data collection, we processed the data to prepare it for ingestion into a machine learning model. We chose to use Autoencoder\footnote{https://blog.keras.io/building-autoencoders-in-keras.html}, because it uses a neural network to encode data into a low dimension and then decode it, attempting to minimize reconstruction loss. It's able to perform anomaly detection by checking the magnitude of the reconstruction loss \cite{adwithauto}. In other words, the Autoencoder's inability to reconstruct particular data implies that the data is anomalous. Using this method we were able to achieve high accuracy and train and predict quickly as elaborated in the later sections. Our success using this model encourages us to test other deep learning models in the future and compare their metrics.

The main contributions of this paper as as follows:
\begin{itemize}
    \item We developed a scalable smart farming environment and collected data from different sensors. We also highlight some challenges faced and solutions we designed.
    \item We designed and injected anomalous scenarios along with some natural anomalies encountered in our smart farming environment.  
    \item We trained and tested an Autoencoder model which is an unsupervised artificial neural network.
    \item We demonstrated how an Autoencoder model can perform well with promising results.  
\end{itemize}

The remainder of the paper is organized as follows: Section \ref{sec:related}  discusses the related works in anomaly detection and security issues in smart farming. We introduced the deployed smart farming architecture in Section \ref{sec:arch} elaborating on the different sensors, hardware and software used to conduct our experiments. Section \ref{sec:data} highlights the entire data collection and processing stages along with the different anomalies injected into the system. Description of the Autoencoder machine learning model and its architecture is done in Section \ref{sec:ml-model}. The results generated by the model are discussed in Section \ref{sec:results} followed by conclusion and future work in Section \ref{sec:conclusion}.

\section{Related Work}
\label{sec:related}
The exponential rise in number of internet connected devices has raised security concerns, especially in the agriculture sector, as farmers will not be able to bear the potential loss and damage to crops. 
Gupta et al. \cite{gupta2020security} developed a comprehensive survey of issues in the security and privacy of the IoT in 2020. The paper covers the architecture of a smart farm environment in-depth and potential real world attack scenarios. It focuses on smart farming in its mature form, in which devices are interconnected with one another and also connect to the internet.  Jeba et al. collected pH and moisture data from soil sensors connected to an Arduino then visualized the data using ThingSpeak. Thingspeak provides a way to visualize and monitor farm conditions through the use of WiFi. In our deployed architecture, we collected data locally using a raspberry pie, acting as a smart edge based system \cite{jeba2018anomaly}. Sontowski et al. \cite{sontowski2020cyber} described various types of network attacks that could be orchestrated on smart farms. The authors also targeted a smart farming test bed with a Denial of Service (DoS) attack. This attack and the others described in the paper are ways an attacker can manipulate or harm the system. The approach presented in our work will be able to identify certain behavior in the form of anomalies and later potentially report the findings to the user. 

A seminal work by Chandola et al. \cite{chandola2009anomaly} titled ``Anomaly Detection: A Survey" offers a comprehensive study of types of data, anomalies, and techniques for detection. It also details all domains in which anomaly detection has been used. This detailed survey provided the authors of this paper with much of their fundamental understanding of the topic. Several papers \cite{park2021anomaly,hasan2019attack, cook2019anomaly, luo2018distributed, chukkapalli2020ontologies, sedjelmaci2016lightweight,kimmell2021analyzing,gupta2021detecting} offered specifics on anomalies in smart ecosystems. In smart farming, the sensors used in a connected farm are often exposed to harsh environments. The conditions make the devices prove to failure, malfunction, attacks, tampering, etc. Any of these conditions could cause abnormal device readings, which would be considered anomalous compared to normal data \cite{gaddam2020detecting}. The low cost of IoT devices in general and the associated implications are further elaborated in later sections of this paper. 
Kotevska et al. \cite{kotevska2019kensor} from Oak Ridge National Laboratory created an algorithm called "Kensor" which aims to achieve detection of normal and abnormal behavior, except instead of individual sensor data they explored co-located/coordinated sensors. These sensors are located in close proximity to one another and the combined collected data from each is used to paint a picture of normal or abnormal. This algorithm offers a solution to more interconnected systems than ours.  However, the inclusion of co-located or coordinated sensors aids the anomaly detection tool in its performance and will be considered in future work. Guo et al. \cite{guo2018multidimensional} proposed a model called "GRU-based Gaussian Mixture VAE system" for anomaly detection in multivariate time-series data. GRU (Gated Recurrent Unit) cells are used to discover correlations among time sequences. Gaussian mixture means that the model combines several Gaussian distributions rather than the more common distribution, Gaussian single-modal. The authors found that this model outperformed traditional Variational Autoencoders (VAEs) in tests on accuracy and F1 score.  *** Yin et al. \cite{8986829} tested a Convolutional Neural Network (CNN) for anomaly detection on time-series data. The authors found that the resulting metrics were promising and achieved a desirable result in anomaly detection. BigClue Analytics \cite{hurubigclue} is a middle-ware solution that offers data approximation, sampling, parallel processing, and anomaly detection in a low-latency scenario. Our focus is not currently in low-latency solutions. However, this would be useful for systems already "online". They tested statistical algorithms, as well as linear regression, signal decomposition, and other methods of anomaly detection. The authors also mention several downfalls of statistical methods for time-series anomaly detection including: missed subtle outliers, inability to detect multiple consecutive outliers, lack of predicted values, etc. In our work, we chose not to use a statistical machine learning method for our system for these and other reasons mentioned in Section \ref{sec:ml-model}. The BigClue Analytics paper chose to use a smart greenhouse as their use case as well. Their interval of sampling was much longer than ours at every 15 minutes, and we believe that this amount of time is too long to wait between samples. They also only evaluated temperature and humidity data, whereas we look at several more sensor outputs. Lastly, ARIMA \cite{zhang2003time} was suggested as a popular time-series analysis technique that we intend to explore in our future work. All of these related works offered solutions that helped us refine our choices involving model selection. We will consider some of the approaches in our future work.

Throughout our literature review, we found that there is an overall lack in studies done on the behavior of specific sensors. We wanted to contribute to the understanding of IoT sensors for smart farming by creating a working environment and collecting a large set of data. We will also be publishing the dataset for public use. Since smart farming is a relatively new domain, we found that efforts to improve the systems and/or detect anomalies were lacking. This is what we hope to offer with this paper and future work.

\section{Experimental Setup, Challenges  Encountered and Solution Approaches}
\label{sec:arch}
\begin{figure*}[t!]
    \centering
    \includegraphics[width=.75\textwidth]{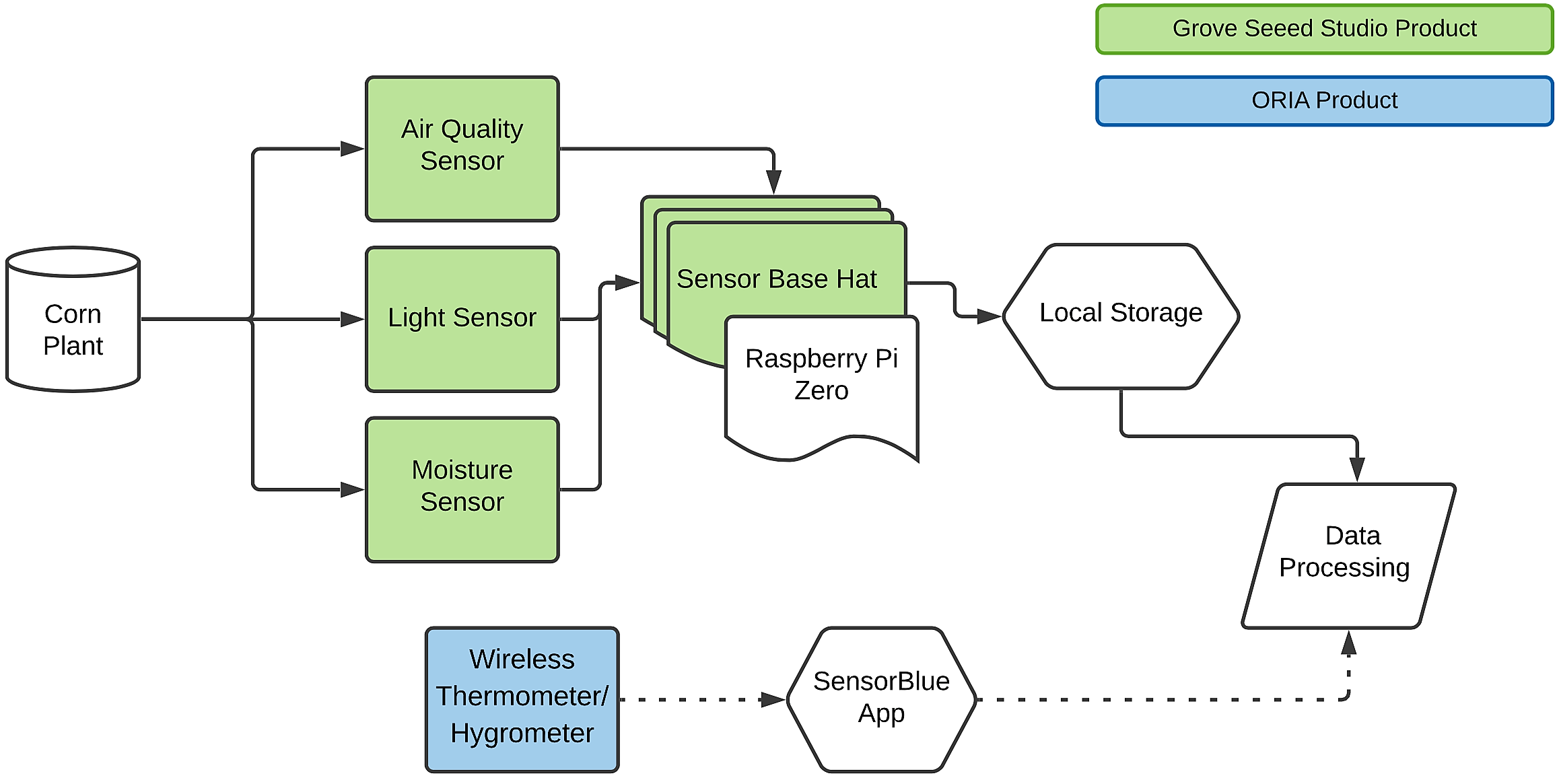}
    \caption{Overview of Deployed Architecture}
    \label{fig:arch}
\end{figure*}

The architecture of our deployed smart farm ecosystem is shown in Figure \ref{fig:arch}. We used Grove Sensors\footnote{https://www.seeedstudio.com/category/Sensor-for-Grove-c-24.html}, a Raspberry Pi Zero\footnote{https://www.raspberrypi.org/products/raspberry-pi-zero/}, a Grove Base Hat to connect the sensors to the Pi, an ORIA Temperature Sensor/Hygrometer, and basic peripheral hardware items including a monitor, keyboard, and mouse. The Grove sensors used were to collect data on air quality (Grove - Air Quality Sensor v1.3), light readings (Grove - Light Sensor), and soil moisture values (Grove - Capacitive Soil Moisture Sensor (Corrosion Resistant)). The ORIA sensor was chosen for its Bluetooth connectivity capability and accompanying smartphone application SensorBlue. The device was placed in our greenhouse environment and when the smartphone with the SensorBlue app was close to the device, it would dump all of the accumulated data to the app via Bluetooth which was then later exported and  formatted into a CSV file. This was helpful in allowing us to select the exact date range that we used for the Grove sensors, and for which we want to process the data. 
The Raspberry Pi Zero was chosen because it is wide availability and cost effective with all the connectivity capabilities needed in a smart farming system. It also works well with the Grove Base Hat which is necessary for connecting to sensors. All of the peripheral hardware we used was for starting and stopping data collection, and later for moving the datasets to our personal machines. 

\begin{figure}[t!]
    \centering
    \includegraphics[width=6cm]{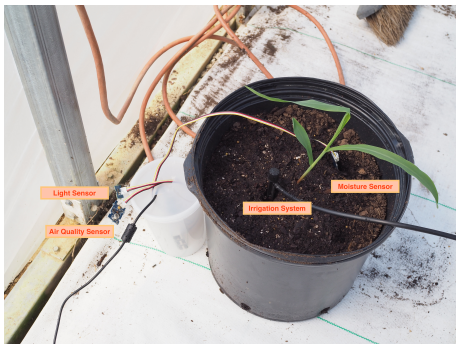}
    \caption{Data Collection Set-Up at Shipley Greenhouse}
    \label{fig:datacoll}
\end{figure}

Initially, the sensors were deployed in a single house plant to fine-tune parameters for data collection such as sampling interval, boundary conditions, and time and date formatting. After collecting data for 10 days indoors, we made changes to the Python source code that controls the sensors. Next we were able to move our sensors to Shipley greenhouse owned by Tennessee Technological University (TTU). This allowed us to collect data in an environment where the temperature, light conditions, and watering schedule of the plant in question were more variable than they were indoors, providing a real-world environment for data collection. Our collaboration with the agricultural department of TTU was instrumental in helping us collect data for our model. Scientists there also provided us with the watering and light \textbf{schedules} so that when we observed the data we would be able to identify patterns of behavior. These schedules in turn helped us determine the types of values that would merit labeling an event as anomalous or not. A photo of the deployment set-up in a plant is shown in Figure \ref{fig:datacoll}. Further expansion of this work could involve placing sets of sensors in several plants and connecting them all to a central Raspberry Pi. This would allow more precise comparisons of data and more accurate anomaly detection.

\subsection{Hardware and Sensors Deployed with Limitations}
Grove sensors, though capable of sensing within our necessary accuracy, are on the low end of the price range, analogous to any smart farming deployment. The sensors we used all cost less than \$10, which are kind of what will be likely used in a real ecosystem. Through extensive periods of exposure to heat and moisture, the sensors were prone to fail. The environment inside the greenhouse was consistently in the 90\degree range. Several times throughout the data collection period, the sensors disconnected from the Raspberry Pi temporarily, stopping the stream of data. In order to prevent this and collect data continuously, the sensors and the Pi were enclosed in a box and stored away from direct sunlight and near air circulation. Unfortunately, this also caused several issues which are discussed below.

Another issue with IoT devices is their interconnectivity. In our case, all of the Grove sensors are connected to the Raspberry Pi through a Grove Base Hat. There are several vulnerabilities associated with this aspect of the system. Each sensor connects to ports on the Grove Base Hat with 4-pin cables. These cables fit into the ports loosely and are easily disconnected. If one sensor's connection is compromised and stops collecting data, and the rest of the sensors continue to collect, all of the data will be associated with different points in time. A benefit to the connectivity between devices is that if each Grove sensor shows the same anomalous pattern, it could be clearer where the anomaly originated, likely in the Grove Base Hat, the Raspberry Pi, or somewhere further upstream. Another physical vulnerability is that we stored the Raspberry Pi and sensor set-up inside a plastic container, as mentioned above. The cables were fitted through the top and then placed on or around the plant we used for the experiment. The sensor cables were prone to unplugging during set-up and we sometimes did not know until we went back to the farm to look at the data we collected. More snug fitting cables could make connectivity more reliable and so could creating a higher quality storage box for the devices.  

It must be noted that the issues highlights in this discussion are specific to our deployment, and we believe in more realistic settings such concerns will be taken care in advance. Our goal in this paper is not to discuss or highlight architectural limitations, but to propose a novel solution to detect anomalies in smart farming ecosystems.

\subsection{Sensor Source Code Modifications}
\label{Subsec:soft}
We deployed the Grove sensors using the Python source code files available with each Grove product. These can be accessed through  GitHub\footnote{https://github.com/Seeed-Studio/grove.py}. We altered several sections of the source code to better suit our system. One of the changes we made to each file was to save the data we collected locally. We decided to store the data this way because the greenhouse we used for our project did not have reliably functioning WiFi, which could be the case in a real smart farm as well. Storing locally gave us a way to have access to the data and to leave the sensors running without worrying that a lost internet connection could be contributing to any discrepancies in the data. Although, attacks using internet connectivity \cite{sontowski2020cyber} and or signal connectivity issues would cause anomalous spikes in the data in a true smart farm environment, we ignore such attacks here in order to gain an understanding of what the collected data should look like with as few interruptions as possible. The data was stored locally as a CSV file to make processing easier later in the project. Along the same lines, we removed any non-numeric data being stored by the sensors. The numeric readings were often originally saved with a descriptor such as ``High Pollution" for the air quality sensor or ``Dry" in the case of the soil moisture sensor. These descriptors could not be processed in our model in any meaningful way so they were removed. We also changed the interval of sampling for each Grove sensor to every 60 seconds. Initially, the sensors were set to collect every tenth of a second. This produced too many data points within the time we allotted for data collection. It also created an issue in matching all the sensor outputs together. Changing the interval to 60 seconds meant that we could group all reading by the minute they were collected. We found no issue in reducing the number of samples in terms of lost specificity among changes in the data. However, the temperature and humidity sensor stores one data point every 10 minutes. It did not have source code that we could access, so we made an alternative change, as elaborated in the data processing subsection \ref{subsec:dp}. The last general change made was to the time and date formatting. We used the \texttt{time.asctime()} function in Python to ensure we could separate data points in the CSV file for processing \cite{pythontime}. Each Grove sensor's source code had individual changes made to them as outlined below. 

\subsubsection{Air Quality Sensor}
For the most part, the sensors' readings are categorized using \texttt{if-else} loops. The air quality sensor has two groups that readings can fall into: ``High Pollution" and ``Air Quality OK". Originally the threshold for a High Pollution reading was a value of 100 or more. The information page\footnote{https://www.seeedstudio.com/Grove-Air-Quality-Sensor-v1-3-Arduino-Compatible.html} for this sensor said it is responsive to a variety of harmful gases. The specific gas that causes the reading cannot be determined, only that the level of the harmful gas is high. We performed several tests to see how easy it would be to reach a 100 reading. We exposed the sensor to a can of wood stain, which contains potent toxic gases. This raised the reading to around 130. This led us to alter the threshold for ``High Pollution" to a lower value, specifically 40, because exposure to harmful gases similar to wood stain would be damaging to the plants. Setting the threshold at 40 allowed us to see when there were small but out-of-the-ordinary changes in the air quality. 

\subsubsection{Light Sensor}
The light sensor contains a photo-resistor which detects changes in light intensity in the environment. The brighter the light is, the lower the resistance of the sensor is. Lower resistance means that a higher voltage can be achieved, making the sensor value high in bright conditions and low or even zero in the dark. We found that our data reflected this well. The sensors read values between 0 and 10 in the hours of the night and up to 640 in the brightest part of the day. The readings produced clear separations between night and day, so we didn't change anything regarding the thresholds for this sensor. 

\subsubsection{Capacitive Moisture Sensor}
As the name suggests, this sensor uses changes in capacitance to determine the level of moisture in soil. The readings are counter-intuitive because the more moisture is detected in the soil, the lower the output reading. The sensor initially had two groupings: ``Wet" and ``Dry". We performed several tests and determined that the readings would be better separated into three groups. Originally the threshold for a ``Dry" reading was a value between 0 and 300 and ``Wet" reading was anything greater than 300. When we started our initial set-up and testing of the sensors, we found that when the sensor was exposed to soil that had just been fully watered, it produced a ``Dry" reading. Several tests were performed to make sure that this was not a fluke. Ultimately we decided we would need to alter the code to switch the values around to achieve consistency in the readings and environment. After changing the values in this way, we also found that the values of 0 up to 600 did not fully encompass what we wanted them to. First, the sensor was placed in the soil approximately 6 inches away from the plant. This meant that when the irrigation system turned on for the plant, the sensor would be exposed to running water. This produced values over 1900. This led us to changed the threshold for a ``wet" reading to anything greater than 1900, ``Moist" to between 1300 and 1900, and ``Dry" to any reading less than 1300. Although ultimately we removed the language descriptors from the data before processing, adding in the third grouping helped us understand the sensor readings better.

\section{Smart Farming Data and Anomalous Behaviour}
\label{sec:data}
 Data produced from Grove Sensors is time-series data. Time series data is a collection of quantities that are assembled over even intervals in time and ordered chronologically \cite{fu2021hawatcher}. Information collected from any ”smart” environment is considered actionable, in that it can be used to make decisions and take actions on critical systems \cite{adwithauto}. This means that the actuation processes are directly affected by all the data in a system. Reducing anomalous patterns and alerting the user when data is out of the ordinary is imperative to keep smart systems running smoothly. The trouble in identifying anomalies in sensor data is that "normal” can look different every day and oftentimes, several times within one day. On a sunny day with high light and temperature readings, you would expect all the data for the day to flow together seamlessly. However, a thunderstorm could suddenly blow through, and cause a decrease in light and temperature readings. This would be considered anomalous, but not in a \textit{harmful} way. On a day with no storms, a quick decrease in the same sensor measurements would be considered anomalous and potentially harmful or incorrect. Detection of anomalies, both harmful and benign, is the goal of this work.

\subsection{Data Collection Phase}
 \begin{figure}[t!]
    \centering
    \includegraphics[width=9cm]{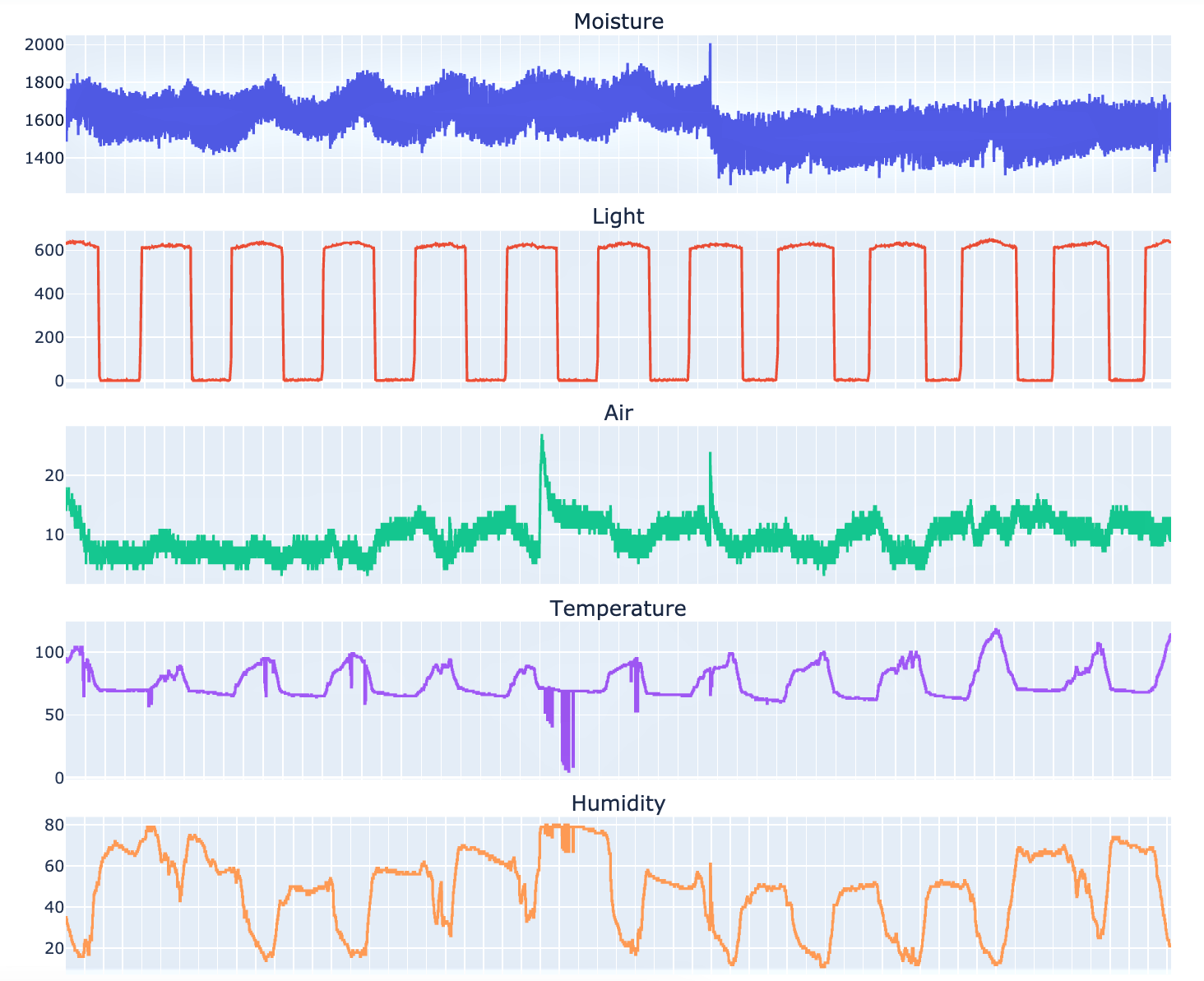}
    \caption{Data Collected Over 11 Non-Consecutive Days}
    \label{fig:data}
\end{figure}

We began data collection at the greenhouse in the first week of April in 2021. We used our understanding of the weather (based on historical data) in Middle Tennessee region during the spring to help us understand what normal readings should be. In the first phase, the sensors collected data for 7 days and were stopped. 
This was the first time all of the sensors had run simultaneously in an outdoor real environment and with the changes we made to the interval of collection in the source code as discussed in previous section. The data was collected every 60 seconds. Roughly 10,100 data points were collected for the first phase which we started processing for our ML model. In month of may, we began the second phase of collection. After processing the first dataset, we decided it would be beneficial to train the machine learning model we chose with more data. This time the sensors ran for 6 days and produced approximately 7200 data points. The total number of data points collected was 17349. We believe this dataset\footnote{Please send an an email to mgupta@tntech.edu, if you are interested in working with the dataset.} can be used by the community interested in learning more about the behavior of Grove and ORIA sensors, and working on smart farming research. 

As shown in Figure \ref{fig:data}, the data translated well visually. The light sensor provided extremely consistent readings, showing distinct night and day times. Other sensors provided less distinct readings, but were generally consistent overall. There are several factors worth mentioning here. For the first 7 days, the moisture sensor fluctuated up and down daily. The change over to the second collection set is distinguished by a sharp peak and then a more level set of points for the next 6 days. As shown in Figure \ref{fig:datacoll}, the moisture sensor is the only one installed in the actual plant. The difference in readings in the first and second set could be due to a different placement in the plant with reference to the irrigation system location. It could also be due to whether the sensor was put in the soil with its electrodes facing the water supply or away from it. It is worth mentioning that the orientation of each of the sensors has the potential to drastically affect the sensor readings. We paid close attention to this after we found the jump in the moisture data.

Creating this environment from scratch meant several things about the data we received. First, this data is unlabeled. It was taken directly from the sensors and then imported for processing. As mentioned in subsection \ref{Subsec:soft}, we made small adjustments to the number of data points collected by each sensor so that they would all match up. However, we did not individually label each data point as anomalous or not. With over 17000 data points, this would have been incredibly time consuming and difficult to achieve accuracy in. \textit{Therefore, the model we selected needed to be able to process unlabeled data, and an Autoencoder is capable of this, as discussed in the next section.} Also, the data falls into the time-series category. Time is an attribute of these data points and implies a correlation between neighboring points. This data is known to contain multiple types of irregularities: contextual, global, and collective \cite{anodot}. Our dataset had predominantly contextual and global outliers. Contextual means that the outlier is only out-of-the ordinary for the time slot it was found in or among its neighboring points. Global means the anomalous point would be anomalous no matter where it was found in time, or with reference to other points. The severe drop in Temperature data in Figure \ref{fig:data} would be an example of both. No other data points were this low in temperature, but even at the lowest temperature, there were no other spikes this low. 

\subsection{Data Processing Phase}
\label{subsec:dp}
In order for our model to properly analyze our input data, we had to pre-process the data before passing it into the model. First, as mentioned in the subsection \ref{Subsec:soft}, the interval of collection between the Grove sensors and the ORIA sensor were different. There was no built-in way to change the ORIA collection interval. To keep the data points separated by a minute, each temperature and humidity reading was copied into each one minute slot encompassed by that ten minute window. For example, on April 16 at 10:19 AM the temperature reading was 85.44 F and the humidity was 36.90\%. This reading was used for each minute from 10:19 AM until the next reading was taken at 10:29 AM. It would be unlikely that a temperature or humidity change large enough to trigger an anomalous reading could occur within these ten minute intervals. Therefore, this is an alteration we felt was relatively low risk. Next, we combined all the data from each of the individual sensors into a single data frame. This helped us visualize data, ensure there were no null values from any sensors, and allowed us to plot our readings more easily \cite{df}.

\subsection{Anomalies}
In order to create anomalies, normal patterns in the data have to be identified first. An anomaly is a pattern or instance in the data that does not conform to a well-defined notion of normal behavior \cite{chandola2009anomaly}. Several anomalous instances occurred naturally in the data. These were removed from the training data set, but left in the testing data set. In Table \ref{table:anoms}, these are listed as "Natural Anomalies". An example of a naturally occurring anomaly is that the temperature sensor has a reading that is "too low". This means that based on a comparison to the current high and low temperatures for Middle Tennessee in April, a reading of less than 54° F would be too low to be normal. It's not so important that this value could never be reached, but that it would be out of the ordinary to get a reading that low. Since we set this floor value ourselves, it would need to be changed based on the location of sampling and time of year. The second section of anomalies is called "Potential Anomalies". These are also bounds on values that we set manually. These statements such as "Moisture too Low" or "Humidity too Low" are readings we would consider anomalous. We selected the bounds for each sensor based on what values we saw over the course of our experimentation. We also injected some anomalies after collection to test the performance of our model. We found that the model was able to detect injected anomalies for all of our sensors. This once again proves that even though we do not train the model with anomalies included in the training data, our approach is still able to detect anomalous readings for each of our devices. A full list of these anomalies can be found in Table \ref{table:anoms}. 

\begin{table}[!t]
    \caption{Anomalies}
    \centering
    \begin{tabular}{| c | c |}
    \hline
    Anomaly & Measurement\\ [0.5ex] 
    \hline
    \multicolumn{2}{|c|}{Natural Anomalies} \\
    \hline
    Temperature too Low & $<$ 54° F \\
    Temperature Difference too Large & $|T1-T2|$ $>$ 25° F \\
    Air Measurement too High & $>$ 20 \\
    \hline
    \multicolumn{2}{|c|}{Potential Anomalies} \\
    \hline
    Moisture too Low & $<$ 1300 \\
    Moisture too High & $>$ 1900 \\
    Light too High & $>$ 640 \\
    Light too Low & $<$ 0 \\
    Air too High & $>$ 40 \\
    Air too Low & $<$ 0 \\
    Temperature too High & $>$ 150° F \\
    Humidity too High & $>$ 90 \\
    Humidity too Low & $<$ 0 \\
    \hline
    \end{tabular}
    \label{table:anoms}
\end{table}

\section{Anomaly Detection Model}
\label{sec:ml-model}
\begin{figure}[!t]
    \centering
    \includegraphics[width=9cm]{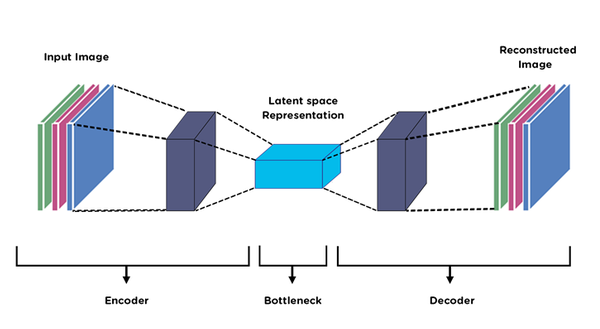}
    \caption{Visual Representation of Autoencoder Model \cite{aediag}}
    \label{fig:modeldiag}
\end{figure}
In this work, our data collection resulted in a massive amount of normal data samples and very few anomalies, therefore it was suitable to train a model with only normal data points and test it with a collection of normal points and the few anomalies that occurred. As such, for anomaly detection, we used Autoencoders \cite{adwithauto} which is an unsupervised neural network technique that learns how to compress and encode data and how to reconstruct the data back from its reduced representation to its original shape. It works by accepting a given input, encoding it into a smaller size using a bottleneck layer, and then decoding it into its original size. Figure \ref{fig:modeldiag} gives a visual representation of an Autoencocer model. It is trained so that the model is able to ignore data that is not crucial to reconstructing the original data as accurately as possible. In addition, the compression feature of Autoencoders helps significantly in dimensionality reduction which makes it capable of processing large number of features, more so than other unsupervised learning methods. 
\begin{figure*}[!t]
    \centering
    \includegraphics[width=.75\linewidth]{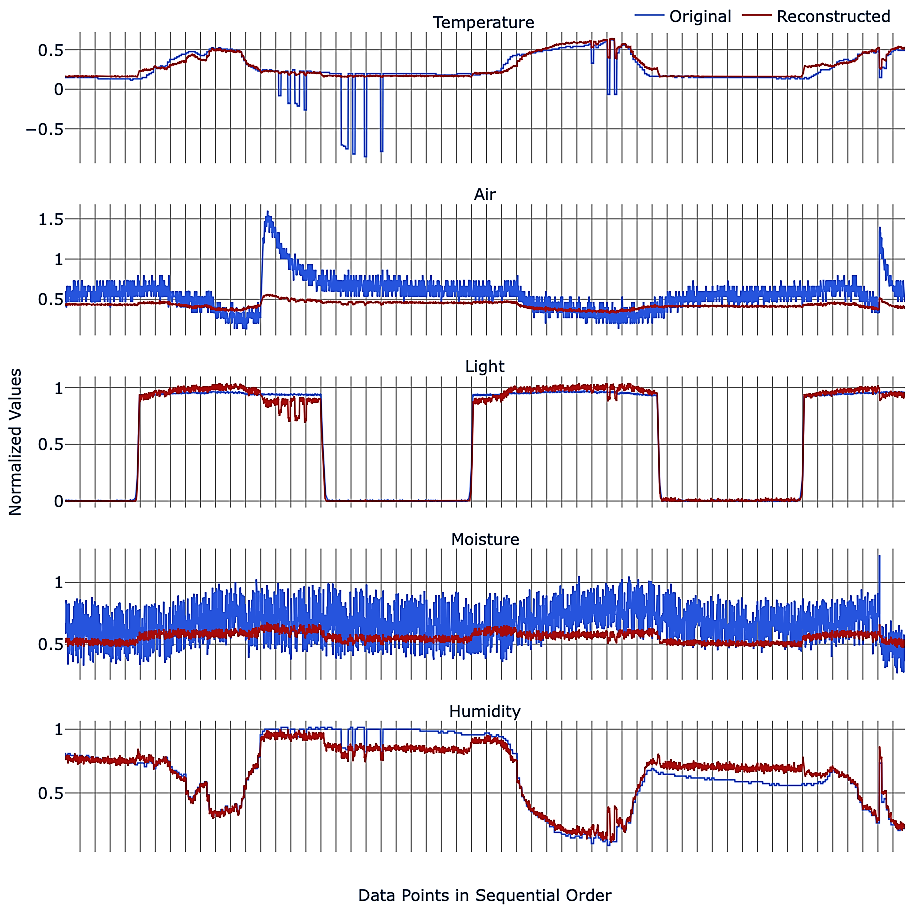}
    \caption{Reconstructed Testing Data Compared to Normal Testing Data}
    \label{fig:recon}
\end{figure*}
In order for the Autoencoder to function properly, it must be trained on data that is scrubbed from any anomalous points exceeding a defined threshold. This ensures that the model is able to reconstruct the normal data with as little reconstruction loss \cite{guo2018multidimensional} as possible, while maximizing the reconstruction loss of the anomalies contained within the testing dataset. Labeling data points as anomalous or benign was achieved by checking each of the data points' attributes for conditions we considered abnormal. The thresholds for this loop can be found in Table \ref{table:anoms}.
The testing data was visually chosen based on a time period that appeared to contain a bulk of anomalous data points, which spanned from April 10\textsuperscript{th} to April 13\textsuperscript{th}.

The values of data points collected from different sensors are of different scales. For instance, the moisture sensor gave measurements ranging from ~1100 to ~2000, whereas the the air quality sensor yielded measurements ranging from ~5 to ~30 (see Figure \ref{fig:data}). As such, the data was normalized using Min-Max normalization \cite{patro2015normalization} to create consistency among the data. Performing normalization on input data has been proven to drastically improve the accuracy of machine learning models. Same parameters used for normalizing the training data were also used to normalize the testing data.


Our model consists of a single input layer, three encoding layers, a single layer for the latent space (aka bottle neck), three layers for decoding, and a single output layer. A diagram showing these basic features is shown in Figure \ref{fig:modeldiag} \cite{aediag}. The input for our model doesn't represent images, but the principle is the same. The model takes an input, encodes it using three layers, then decodes it, and produces an output image. The hope is that by inputting normal data, the model is able to learn what normal behavior is, and recreate it. After the model is trained, we must determine a classification threshold. The hyper parameters such as learning rate, batch size, number of nodes and epochs were determined by conducting a grid search of various values of these parameters until the combination that produced the most optimal results was found The model was trained for 60 epochs, with a batch size of 8, a learning rate of .000001, and a node size of 256.

For model training, the normal data is split into \textit{train} and \textit{validation} data, 75\%/25\% respectively. As mentioned earlier, we have a separate data set specifically for testing. The model was trained using the training data, and the validation data was used to determine the classification threshold which is discussed later on in this section. Our model was trained in 262 seconds.


\begin{figure*}[!t]
    \centering
    \includegraphics[width=.995\linewidth]{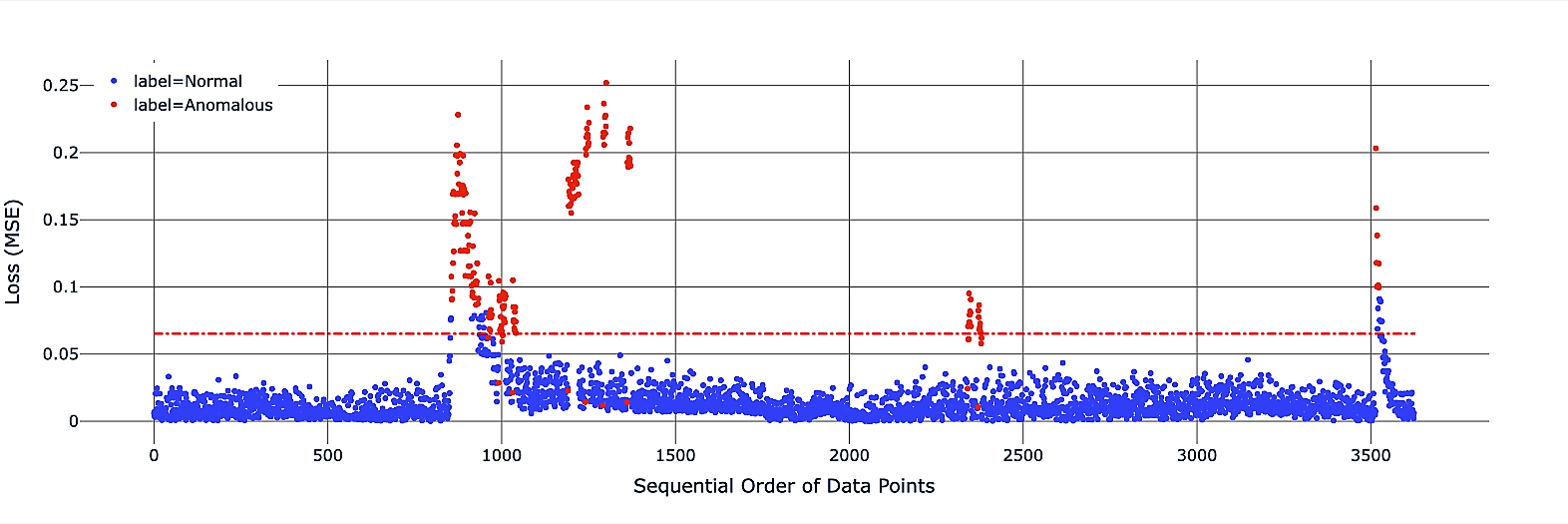}
    \caption{Reconstruction Loss on Test Data}
    \label{fig:thresh}
\end{figure*}
Unlike normal machine learning models that generate predictions, passing data through the trained model does not generate an actual prediction, and instead, it simply returns the loss between the original data point, and the reconstructed data point generated by the Autoencoder model. In order to obtain an actual prediction, we take the loss returned from the model and compare it to a predetermined threshold in order to determine if the data point is anomalous or not. We used our validation data, which contains no anomalous data points, to determine the threshold by taking the average of the 5 highest loss values that were generated. The average was taken in order to prevent one abnormal loss value from skewing threshold while also ensuring that the threshold is larger than the loss generated by a normal data point. The testing dataset is used to evaluate the performance of our model, where it is used as an input to obtain its reconstruction loss. The loss values are then checked against the predetermined threshold to be classified as normal or anomalous.


\section{Experimental Results And Discussion}
\label{sec:results}
The results presented in this section test the performance of the Autoencoder model. We evaluate our model using the performance metrics: accuracy, precision, recall, and F1 score, defined as follow: 

\vspace{-5mm}
\begin{align*}
    Accuracy &= \frac{TP+TN}{TP+TN+FP+FN}
\end{align*}
\vspace{-3mm}
\begin{align*}
    Precision &= \frac{TP}{TP+FP}
\end{align*}
\vspace{-3mm}
\begin{align*}
    Recall &= \frac{TP}{TP+FN}
\end{align*}
\vspace{-3mm}
\begin{align*}
    F1 ~Score &= 2 \times \frac{Precision \times Recall}{Precision + Recall}
\end{align*}

In our experiments, a \textit{positive} outcome means an abnormal activity was detected, whereas a negative outcome means a normal activity was detected.
True Positive (TP) refers to an abnormal activity that was correctly classified as abnormal. 
True Negative (TN) refers to a normal activity that was correctly classified as normal.
False Positive (FP) refers to a normal activity that was misclassified as abnormal.
and False Negative (FN) refers to an abnormal activity that was misclassified as normal.

The success of our model is based on measuring the reconstruction error that is produced by any given data point. Figure \ref{fig:recon} shows an example of reconstructed data overlaid the original data that was inserted into the model. 
In this figure, extremely severe dips in temperature denoted by the blue line (representing our original data) can be noticed. The data reconstructed by the model, represented by the red line, does not dip as much as the original data. This is because our model was not able to reconstruct these points accurately due to the fact that they are anomalies. The reconstruction loss (i.e. different between the original and the reconstructed data), where the model recognizes normal or abnormal behavior, is shown in Figure \ref{fig:thresh}. The figure shows a visualization of the mean-squared-error (MSE) generated by the model after it was given each data point within the test data set. The dotted red line denotes the threshold determined as mentioned in Section \ref{sec:ml-model}. Each data point's actual label is represented either by blue color to denote a normal behavior or red color to denote an anomaly and every data point that lies above the threshold was classified as anomalous. This figure illustrates our model's capability to detect the majority of anomalies by measuring the MSE produced by each data point.

\begin{figure}[t!]
    \centering
    \includegraphics[width=8cm]{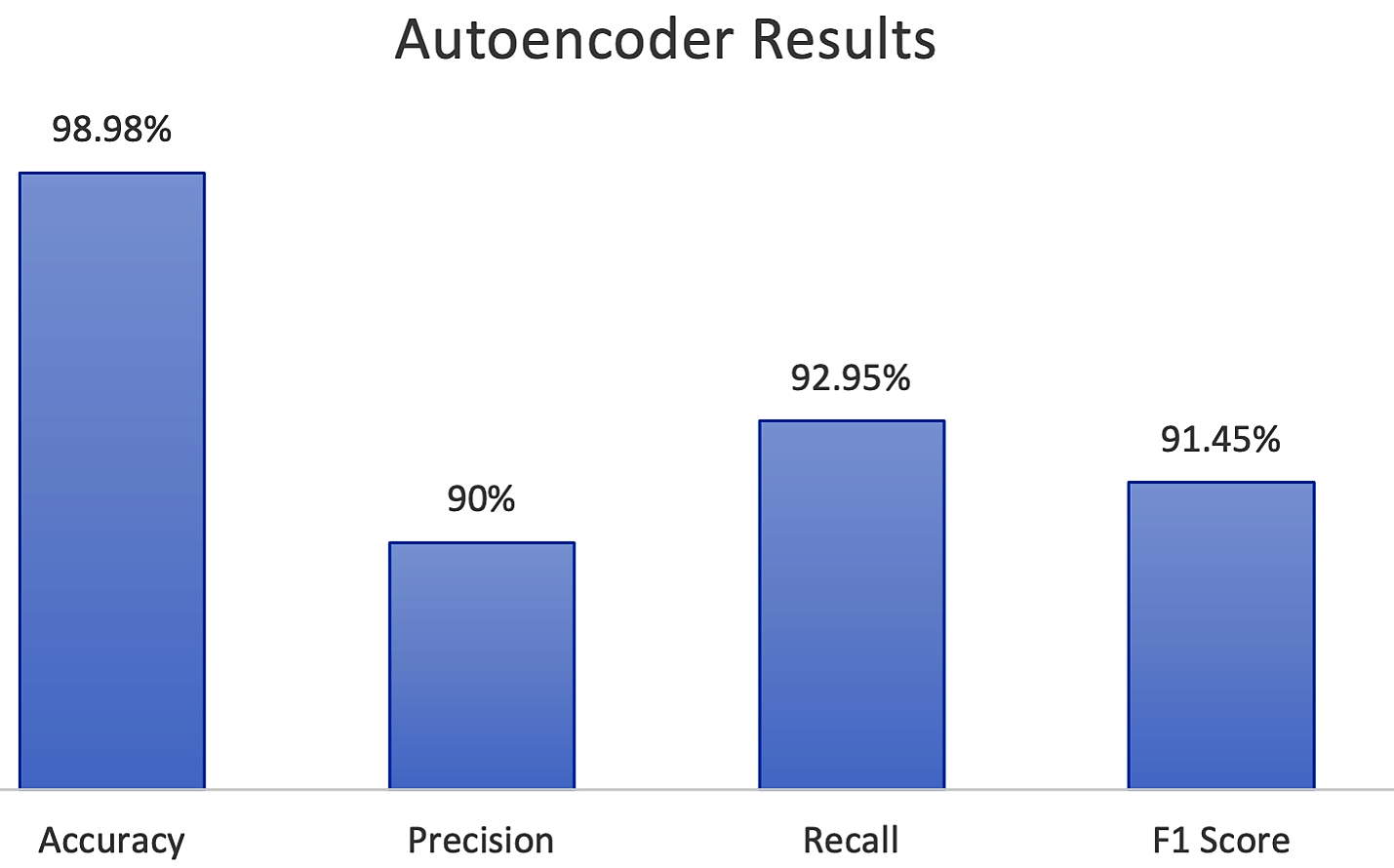}
    \caption{Performance metrics for Autoencoder Model}
    \label{fig:aeresults}
\end{figure}

Overall, as shown in Figure \ref{fig:aeresults}, our model was able to attain high performance with over $90\%$ in all metrics. The precision is lower than the recall metric which shows that the model produced slightly more false positives than false negatives. In a smart farming environment, a higher rate of false positives would not have a dramatic affect on the productivity of day to day operations and would ensure a higher number of anomalous situations are detected. A rather problematic situation would be if there were more false negatives than positives. A user would much prefer receiving an alert when nothing was wrong than not receiving an alert and enabling potential harm to occur to the crops and hardware. In the future, we hope to further decrease the number of false positives and negatives in order to fine-tune an overall more accurate model. This can be done by using more training samples.

    
    

\section{Conclusion and Future Work}
\label{sec:conclusion}
Our approach has shown that smart farming anomaly detection can be done at an extremely accurate level by using an Autoencoder. Our approach would allow vast scalability by only requiring non-anomalous data for training. Greenhouses provide controlled environments that create consistent conditions for crops and data collection. Environments such as this are a perfect use case for our approach since the performance of an Autoencoder can drastically improve when provided with large amounts of non-anomalous data. Our approach shows that it may not be entirely necessary for machine learning professionals that are working on anomaly detection within smart farming to be highly concerned with developing models that are trained using labeled data that contains both normal and anomalous data. 

In the future, we will explore more anomaly detection models in order to optimize the system's performance. Once the best model has been selected, the architecture could be brought online to be used and tested with the added interactions of Internet connectivity. By bringing the system online we will have the ability to alert users of potential threats or anomalous behavior. These alerts could be coupled with actuators such as fertilization, watering, video monitoring, etc. The introduction of cameras can be ``used to calculate biomass development and fertilization status of crops" \cite{Walter6148}. They can also be used to allow the system-user to monitor their property from afar. We plan to introduce photo and video monitoring as one of our next steps to improve security and broaden our scope.

\section{Acknowledgements}
\label{sec:ack}
We thank TTU Shipley Farms for allowing to use greenhouse, and setup smart farm testbed. Dr. Brian Leckie and his group were instrumental in our system and early stages of data collection. We are thankful to Ms. Deepti Gupta to provide helpful guidance on dealing with time-series, correlated data and gave input on our model selection. This research is partially supported by the NSF Grant 2025682 at TTU.

\bibliographystyle{./bibliography/IEEEtran}
\bibliography{./bibliography.bib}

\vspace{12pt}

\end{document}